\documentclass[conference]{IEEEtran}
\IEEEoverridecommandlockouts
\usepackage{cite}
\usepackage{amsmath,amssymb,amsfonts}
\usepackage{algorithmic}
\usepackage{graphicx}
\usepackage{textcomp}
\usepackage{xcolor}
\usepackage[a4paper, total={184mm,239mm}]{geometry}
\def\BibTeX{{\rm B\kern-.05em{\sc i\kern-.025em b}\kern-.08em
    T\kern-.1667em\lower.7ex\hbox{E}\kern-.125emX}}
\begin{document}

\title{SEGA-DCIM: Design \underline{S}pace \underline{E}xploration-\underline{G}uided \underline{A}utomatic \underline{D}igital \underline{CIM} Compiler with Multiple Precision Support
\thanks{$^{*}$These authors contributed equally to this work. This work was supported in part by the Beijing Natural Science Foundation (L244051), the Beijing Major Science and Technology Project (No. Z241100004224004), the 111 Project (B18001), and the Beijing Outstanding Young Scientist Program (JWZQ20240101004). Corresponding author: Xiyuan Tang.}
}

\author{\IEEEauthorblockN{Haikang Diao$^{*1}$, Haoyi Zhang$^{*1}$, Jiahao Song$^2$, Haoyang Luo$^{1}$, Yibo Lin$^1$, Runsheng Wang$^1$,\\
Yuan Wang$^1$, Xiyuan Tang$^{1}$}
\IEEEauthorblockA{
$^1$Peking University, Beijing, China ~~$^2$University of California at San Diego, CA, USA\\
Email: diaohaikang@stu.pku.edu.cn, xitang@pku.edu.cn}
}

\maketitle

\begin{abstract}
   Digital computing-in-memory (DCIM) has been a popular solution for addressing the memory wall problem in recent years. However, the DCIM design still heavily relies on manual efforts, and the optimization of DCIM is often based on human experience. These disadvantages limit the time to market while increasing the design difficulty of DCIMs. This work proposes a design space exploration-guided automatic DCIM compiler (SEGA-DCIM) with multiple precision support, including integer and floating-point data precision operations. SEGA-DCIM can automatically generate netlists and layouts of DCIM designs by leveraging a template-based method. With a multi-objective genetic algorithm (MOGA)-based design space explorer, SEGA-DCIM can easily select appropriate DCIM designs for a specific application considering the trade-offs among area, power, and delay. As demonstrated by the experimental results, 
   SEGA-DCIM offers solutions with wide design space, including integer and floating-point precision designs, while maintaining competitive performance compared to state-of-the-art (SOTA) DCIMs.
\end{abstract}

\begin{IEEEkeywords}
Digital Computing-in-memory, design space exploration, compiler
\end{IEEEkeywords}

\section{Introduction}
\label{sec:Introduction}

\begin{figure}[tb]
    \centering
    \includegraphics[width=0.35\textwidth]{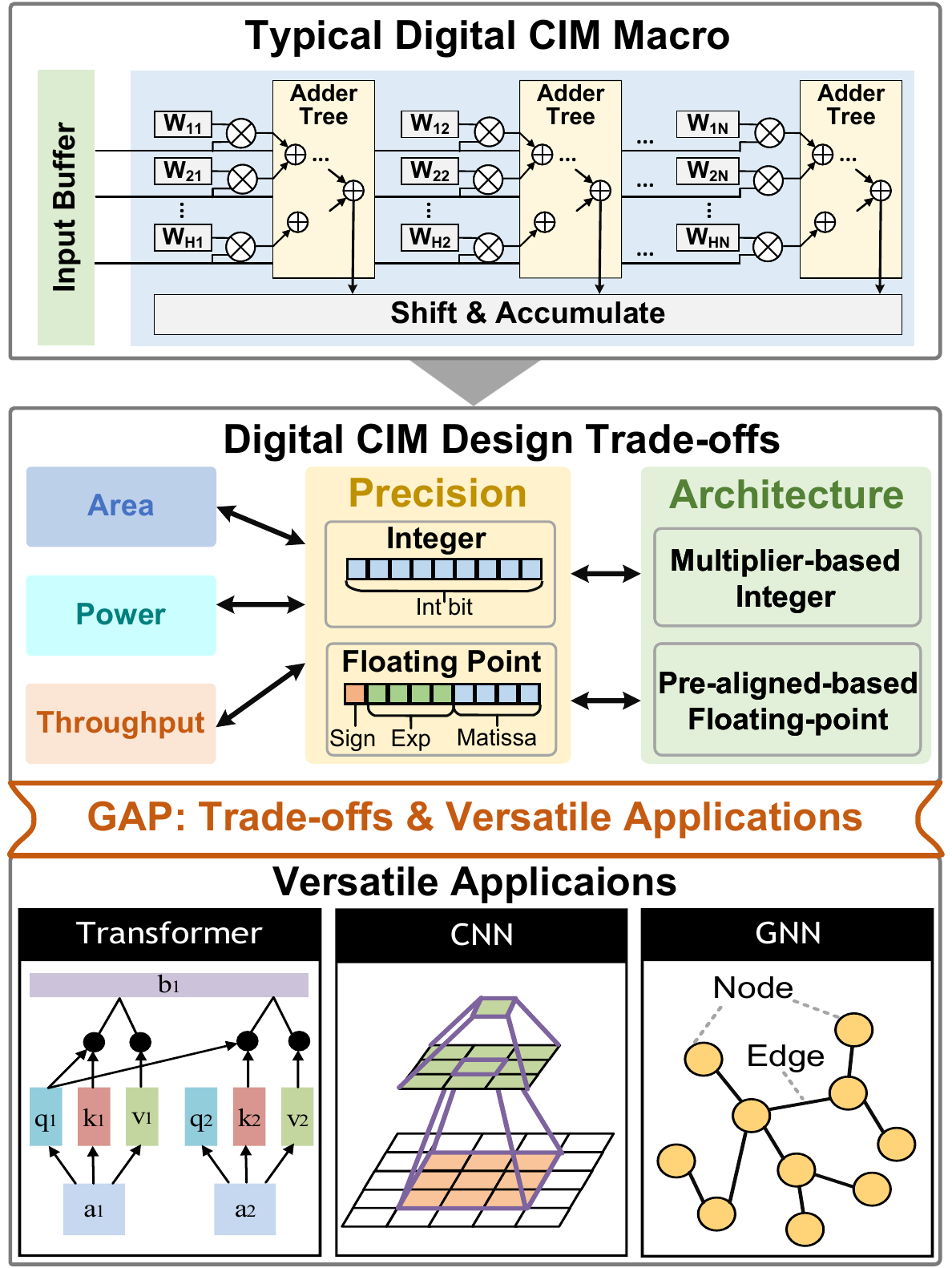}
    \caption{DCIM macro design trade-offs and various scenarios.}
    \label{Intro}
    \vspace{-0.8cm}
\end{figure}

\begin{sloppypar}
Computing-in-Memory (CIM) is a prevailing solution for AI accelerators to overcome the memory-wall problem. It incorporates the compute unit into memory cells to perform multiply-accumulate (MAC) operations locally, dramatically reducing the weight movement's cost. The mainstream CIM designs can be categorized into two types, DCIM and Analog CIM~(ACIM). ACIM has high energy efficiency but suffers from low computational accuracy~\cite{Jia-JSSC, Song-eDRAMCIM}. DCIM uses digital logic circuits to perform the multiply-accumulate computation to achieve full-precision operation, which dramatically improves the reliability and accuracy~\cite{sunAnalogDigitalInmemory, Wang-PositCIM}. In addition, DCIM offers improved flexibility and scalability for expansion to larger array sizes and integration with other digital circuit modules, thus being able to adapt to different application requirements.
\end{sloppypar}

\begin{sloppypar}
Although the flexibility of DCIM is very attractive, this also leads to an expansive design space, complicating the design trade-off of DCIM designs.
As Figure~\ref{Intro} demonstrates, many trade-offs should be considered in DCIM designs when facing versatile applications. The area, power consumption, and throughput are the fundamental concerns of the DCIM. Moreover, as DCIM design continues to evolve, data precision is gradually extended from integer~\cite{dcim1, jssc-hkdiao} to floating-point (FP)~\cite{dcim_float2, Yue_FPCIM, dcim_float3} for high-precision tasks such as model training. Thus, data precision and corresponding architecture should also be considered. Compared to integer data, floating-point data contains exponent bits, significantly increasing the complexity of multiply-add computation. This leads to a more complex architectural design for DCIM and a complete reliance on human experience. Therefore, there is an urgent need for an automated tool to bridge the gap between challenging DCIM design trade-offs and versatile applications. 
\end{sloppypar}

\begin{sloppypar}
In recent years, electronic design automation (EDA) technology has improved dramatically, and some impressive automation tools for CIM have emerged. NeuroSim~\cite{chenNeuroSimCircuitLevelMacro2018,neurosim} tries to benchmark CIMs more efficiently, and several studies~\cite{gonugond2020} attempt to better model CIMs. Furthermore, inspired by end-to-end automated flow for SRAM~\cite{kamineniMemGenOpenSourceFramework2021} and SAR-ADC~\cite{liuOpenSAROpenSource2021}, some research started to automate the CIM design flow. EasyACIM~\cite{zhangEasyACIM} proposes an end-to-end automated flow for ACIMs assisted by design space exploration, which can determine the trade-offs automatically. AutoDCIM~\cite{chenAutoDCIMAutomatedDigital}
proposes an end-to-end automated flow for DCIM, but lacks automated design space exploration and only supports a single architecture for integer precision. The complicated decision of trade-offs is left to the users. Therefore, to reduce DCIM design time and improve design quality, an automated tool that supports multiple precision and automatically determines complicated trade-offs is needed.  
\end{sloppypar}

\begin{sloppypar}
In this work, we propose SEGA-DCIM, a design space exploration-guided automatic DCIM compiler supporting multiple precision. As Table~\ref{tab:comparison} shows, to bridge the gap between challenging DCIM design trade-offs and versatile applications, SEGA-DCIM constructs estimation models for multiple precision operations and introduces a MOGA-based design space explorer to explore the Pareto frontier with a given application automatically. This approach can further improve the design quality and efficiency which is also not considered in AutoDCIM~\cite{chenAutoDCIMAutomatedDigital}. Moreover, SEGA-DCIM leverages a template-based method to generate DCIM designs and commercial tools for digital circuits to generate DCIM layouts. The main contributions of this paper can be summarized as follows: 

\begin{itemize}
    \item We propose an automatic DCIM compiler SEGA-DCIM that supports multiple data precision operations, including INT and FP, which leads to a wider design space. 
    \item SEGA-DCIM constructs concise estimation models for multiple precision, formulates the determination of DCIM trade-offs as a multi-objective optimization problem, and obtains the Pareto frontier by a MOGA-based (NGSA-II) design space explorer. 
    \item SEGA-DCIM integrates a template-based DCIM generator and leverages commercial tools to generate the final layouts according to the optimized design parameters given by the design space explorer. 
    \item SEGA-DCIM generates Pareto-front DCIM designs for a given application scenario with a competitive performance compared to SOTA DCIMs. SEGA-DCIM can be easily extended to new DCIM structures thanks to its template-based approach.
\end{itemize}
\end{sloppypar}

\begin{sloppypar}
The rest of the paper is organized as follows. 
Section~\ref{sec:Preliminary} describes the basic background; 
Section~\ref{sec:Algorithm} explains the detailed implementation; 
Section~\ref{sec:Results} shows the final results; 
Section~\ref{sec:Conclusion} concludes the paper. 
\end{sloppypar}

\section{Preliminaries}
\label{sec:Preliminary}

\begin{table}[tb]
    \centering\
    \label{tab:comparison}
    \caption{Comparison with Other CIM Design Flow.}
    \resizebox{0.5\textwidth}{!}{
        \begin{tabular}{c|ccc}
        \hline
            Entry & EasyACIM~\cite{zhangEasyACIM} & AutoDCIM~\cite{chenAutoDCIMAutomatedDigital} & SEGA-DCIM \\ \hline
            Design type & Analog & Digital & Digital \\ 
            Support precision & INT & INT & \textbf{INT \& Float} \\ 
            Estimation model & Yes & No & \textbf{Yes} \\ 
            Design space & Pareto frontier & Unoptimized & \textbf{Pareto frontier} \\ \hline
            Determination of trade-offs & Automatic & User-defined & \textbf{Automatic} \\ \hline
        \end{tabular}
    }
    \vspace{-0.5cm}
\end{table}

\subsection{DCIM compute architecture}

\begin{figure}[tb]
    \centering
    \includegraphics[width=0.43\textwidth]{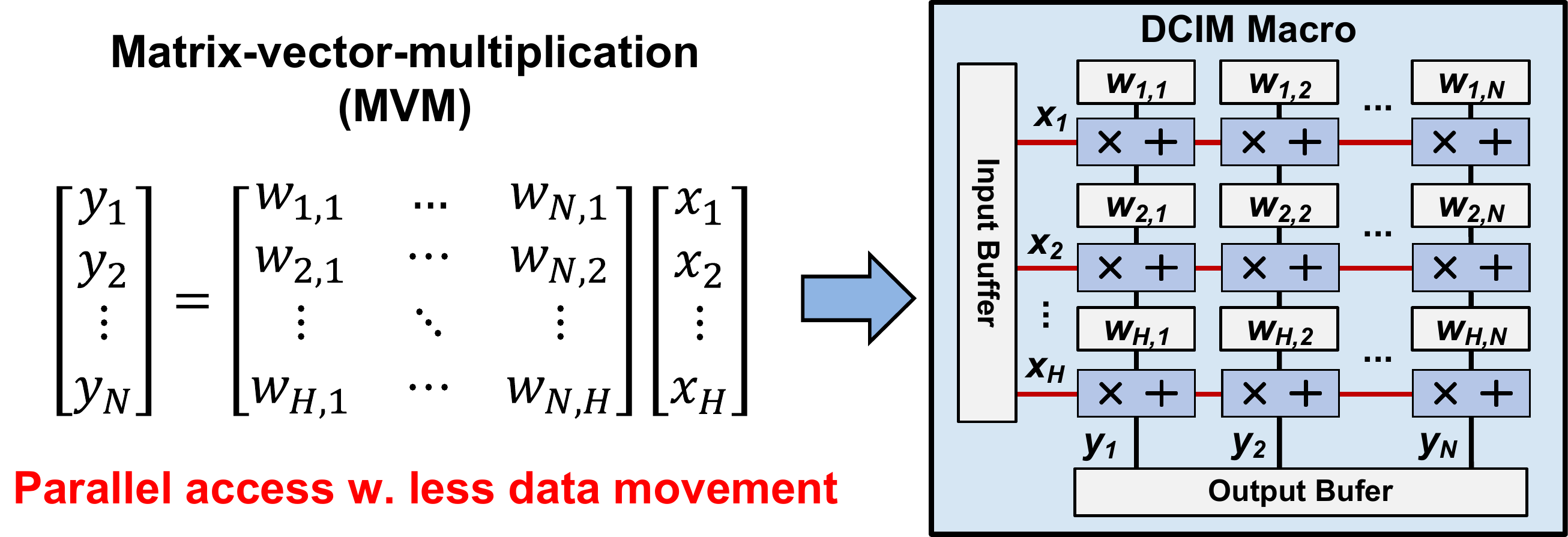}
    \caption{DCIM for matrix-vector multiplication.}
    \label{fig2}
    \vspace{-0.6cm}
\end{figure}

With the increasing number of parameters in the NN model, the power and latency overhead of memory access is much higher than that of computation. The DCIM architecture, by integrating computational units within the memory array, significantly reduces the overhead of memory access, thereby markedly enhancing the computational efficiency of MVMs. As shown in Fig.~\ref{fig2}, the DCIM pre-stores model weights in the memory array. Then, the input buffer continuously feeds input vectors X to the DCIM, enabling parallel computation within the memory array. Upon completion of the computation, the result vector Y can be directly extracted from the output buffer. Throughout the computation, the weights remain static to minimize memory access, leading to a significant reduction in power consumption and latency.

\begin{figure*}[tb]
    \centering
    \includegraphics[width=0.83\textwidth]{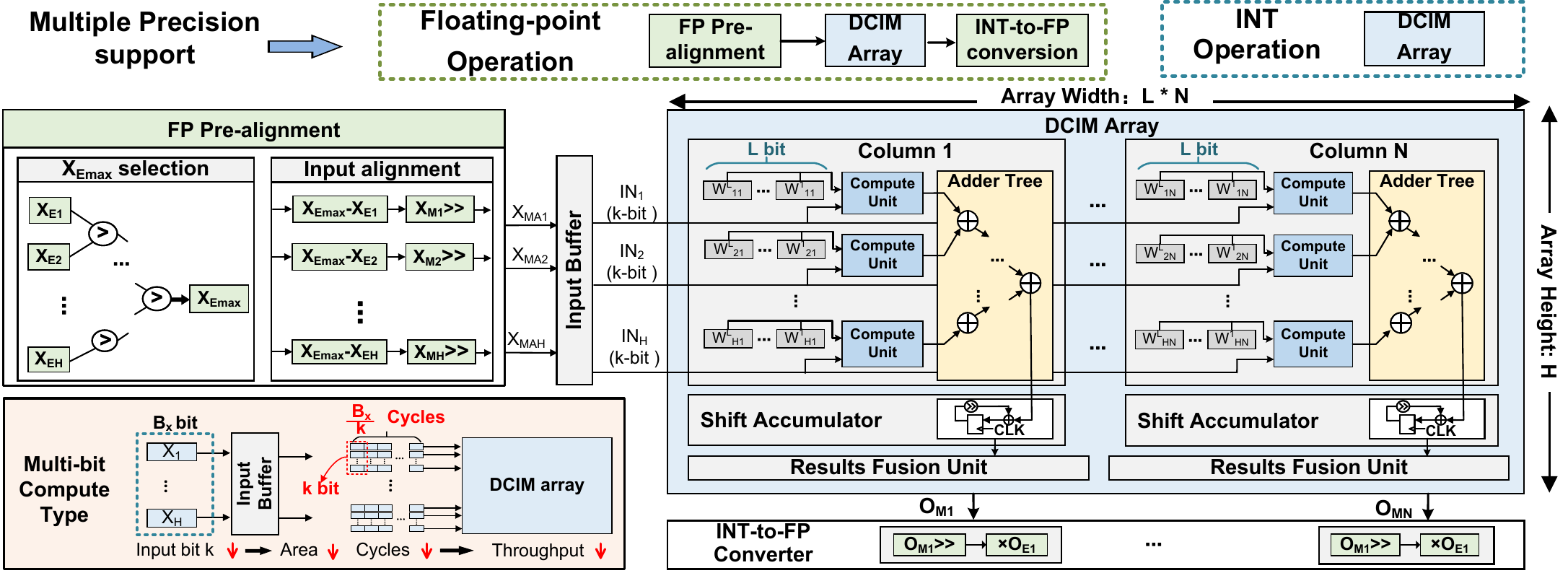}
    \caption{The synthesizable DCIM architecture.}
    \label{CIMArchitecture}
    \vspace{-0.4cm}
\end{figure*}

\begin{figure}[b]
    \centering
    \vspace{-0.6cm}
    \includegraphics[width=0.32\textwidth]{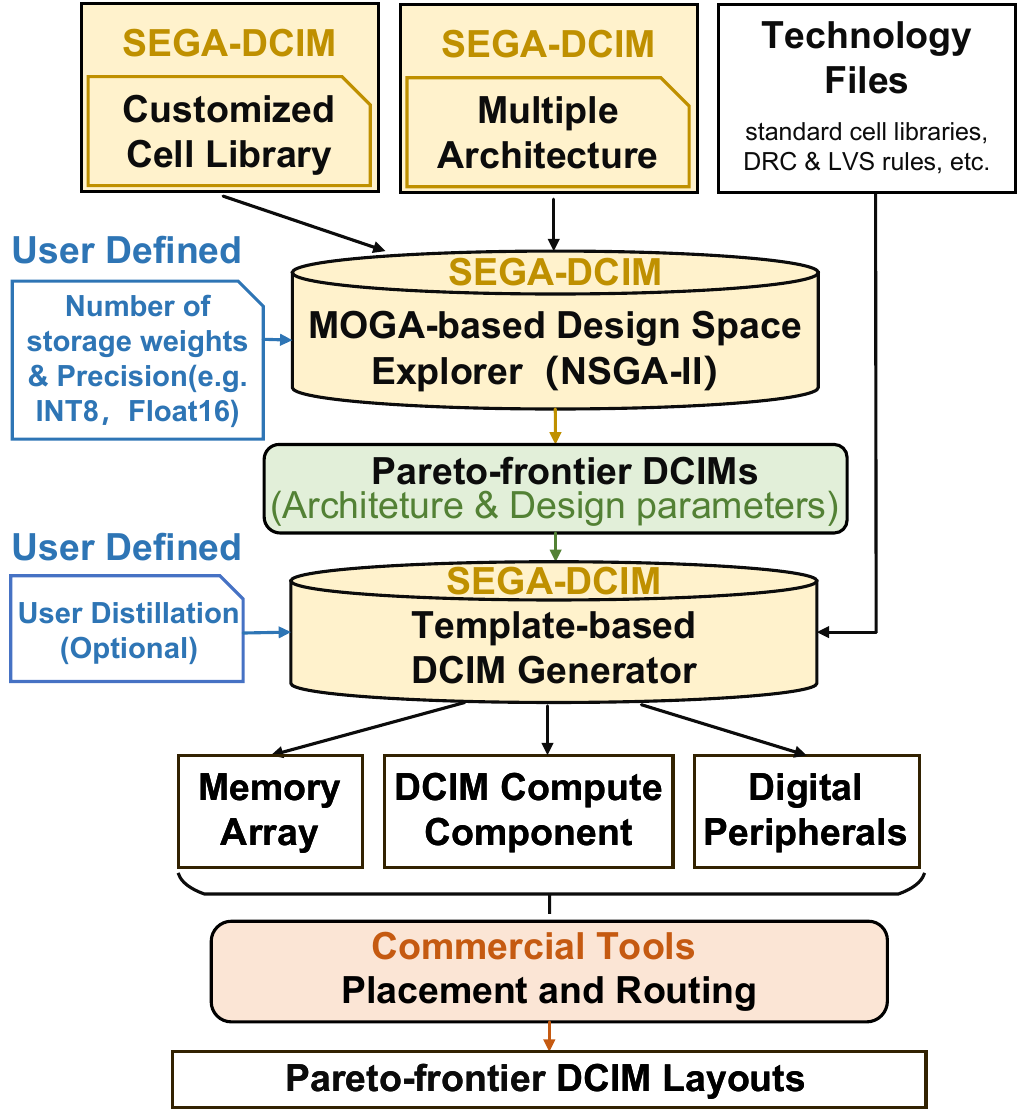}
    \caption{Overview of SEGA-DCIM framework.}
    \label{SEGA-DCIMFrame}
    \vspace{-0.5cm}
\end{figure}

\subsection{Multi-Objective Optimization}
Similar to normal EDA problems such as layout and routing, design space exploration for DCIM can be formulated as a multi-objective optimization problem~\cite{pereiraReviewMultiobjectiveOptimization2022,sainiMultiobjectiveOptimizationTechniques2021}. To simplify the solution process, many previous studies have transformed these multi-objective optimization problems into single-objective optimization problems. However, single-objective optimization often introduces a fixed human experience that is not suitable for multiple architectures and versatile user requirements. Pareto frontier and Pareto dominance are commonly used to compare the candidate solutions of multi-objective problems. Pareto frontier is a group of solutions in which there is no other solution that can improve at least one of the objectives without degrading any other objective. Formally, a solution vector $\vec{u}=\left[u_{1}, u_{2}, \ldots, u_{P}\right]^{T}$ is said to pareto-dominate~\cite{ngatchouParetoMultiObjective2005} the solution vector $\vec{v}=\left[v_{1}, v_{2}, \ldots, v_{P}\right]^{T}$, in a minimization context, if and only if:   

\begin{equation}
    \begin{array}{l}
        \forall i \in\{1, \ldots, N\}, f_{i}(\vec{u}) \leq f_{i}(\vec{v}) \\
        \text { and } \exists j \in\{1, \ldots, N\}: f_{j}(\vec{u})<f_{j}(\vec{v})
        \end{array}
    \label{ParetoFronte}
    \vspace{0.1cm}
\end{equation}

\section{SEGA-DCIM Framework}
\label{sec:Algorithm}

\subsection{Synthesizable DCIM Architecture}

The framework of SEGA-DCIM is shown in Fig.~\ref{SEGA-DCIMFrame}. The framework inputs include the customized cell library, multiple architectures, and technology files. Customized cell library indicates the basic components of DCIMs in the template-based DCIM generation. Multiple architectures include the multiply-based integer architecture and pre-aligned-based floating-point architecture, which are different templates in template-based DCIM generation . Technology files consist of basic information of a PDK such as standard cell libraries, DRC rules, LVS rules, etc. After these inputs are processed, the MOGA-based design space explorer will generate the Pareto-frontier DCIM designs according to the user specifications. Users can give the number of weights, data precision, and any other requirements according to their applications. Specifically, the design space explorer is based on the prevailing genetic algorithm, NSGA-II. Each DCIM design in the Pareto frontier solutions contains the architecture type and corresponding design parameters. After obtaining the DCIM designs based on automatic search, the users can further select their preferred DCIM designs before the time-consuming generation step starts. 

The template-based DCIM generator will only be performed on the selected Pareto frontier solutions. The generation can be roughly categorized into three parts: memory array, DCIM compute component, and other digital peripherals. The netlist generation is based on a customized script, and layout generation leverages popular commercial tools (Innovus). Finally, the commercial tools will finish the layouts for the selected Pareto-frontier DCIM designs.

Fig.~\ref{CIMArchitecture} illustrates the overviews of the proposed synthesizable DCIM architecture, which is designed based on state-of-the-art work. It consists of an FP Pre-alignment, an Input Buffer, a CIM array, and an INT-to-FP converter. When the data format is an integer, the input skips FP Pre-alignment and INT-to-FP converter and is fed directly into the CIM array to calculate the result. When the data format is floating-point, the weight's mantissa is offline aligned and pre-stored in the DCIM array before NN computation. When performing the NN computation, the inputs’ mantissa needs to be aligned in real time. The maximum value of the inputs’ exponent $X_{Emax}$ is first found with a comparison tree. Then, each input’s exponent offset is computed by $X_{Emax}-X_{E}$. Each input’s mantissa is shifted by the exponent offset to achieve alignment. After the mantissa pre-alignment, the aligned input mantissa $X_{MA}$ and weight mantissa $W$ can directly perform INT mantissa MAC in the DCIM array. 

The Input Buffer is used to buffer the aligned mantissa and send $(H*k)$-bits per cycle to the DCIM array for INT mantissa MAC. As shown in the lower left corner of Fig.~\ref{CIMArchitecture}, it takes $\frac{B_{x}}{k}$ cycles to complete the computation for a $B_{x}$-bit mantissa. The smaller $k$ is, the smaller the area of digital circuits in the DCIM array. However, the number of computation cycles $\frac{B_{x}}{k}$ increases, which in turn reduces the throughput.
\begin{figure}[tb]
    \centering
    \includegraphics[width=0.38\textwidth]{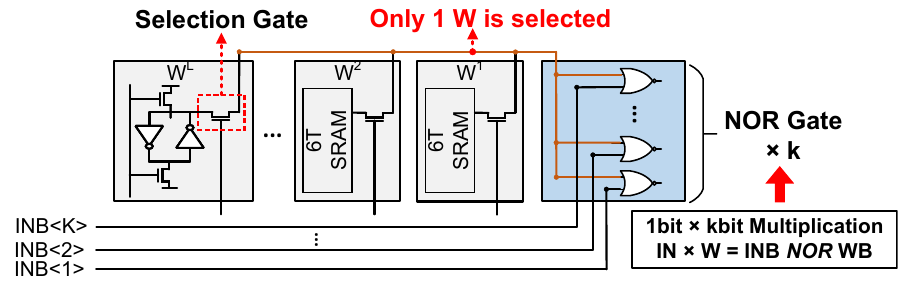}
    \caption{Compute Unit design for Multiply-based DCIM.}
    \label{Multiply-based}
    \vspace{-0.6cm}
\end{figure}

\begin{table*}
\centering
\caption{Symbols and cost model for digital logic modules}
\resizebox{0.9\textwidth}{!}{
    \begin{tabular}{|c|c|c|c|}
    \hline
        \textbf{} & \textbf{Area} & \textbf{Delay} & \textbf{Power } \\ \hline
        \textbf{1-bit*N-bit Multiplier} & $A_{mul}(N)=N\ast A_{NOR}$ & $D_{mul}\left(N\right)=D_{NOR}$ & $E_{mul}\left(N\right)=N\ast E_{NOR}$ \\ \hline
        \textbf{N-bit Adder} & $A_{add}\left(N\right)=(N-1)\ast A_{FA}+A_{HA}$ & $D_{add}\left(N\right)=(N-1)\ast D_{FA}+D_{HA}$ & $E_{add}\left(N\right)=(N-1)\ast E_{FA}+E_{HA}$  \\ \hline
        \textbf{N:1 MUX} & $A_{sel}\left(N\right)=(N-1)\ast A_{MUX}$ & $D_{sel}\left(N\right)=(\log_2{N)}\ast D_{MUX}$ & $E_{sel}\left(N\right)=(N-1)\ast E_{MUX}$  \\ \hline
        \textbf{N-bit Shifter} & $A_{shift}\left(N\right)=N\ast A_{sel}(N)$ & $D_{shift}\left(N\right)=(\log_2{N)}\ast D_{sel}(N)$ & $E_{shift}\left(N\right)=N\ast E_{sel}(N)$  \\ \hline
        \textbf{N-bit Comparator} & $A_{comp}\left(N\right)=A_{add}\left(N\right)$ & $D_{comp}\left(N\right)=D_{add}\left(N\right)$ & $E_{comp}\left(N\right)=E_{add}\left(N\right)$  \\ \hline
    \end{tabular}
    \label{module cost}
    \vspace{-0.5cm}
    }
\end{table*}

\begin{table}[!ht]
\centering
\vspace{-0.1cm}
\caption{Symbols and cost model for standard cells}
\resizebox{0.48\textwidth}{!}{
    \begin{tabular}{|c|c|c|c|}
    \hline
        \textbf{Cell} & \textbf{Area} & \textbf{Delay} & \textbf{Power } \\ \hline
        \textbf{NOR} & $A_{NOR}$=$A_{gate}$ & $D_{NOR}$=$D_{gate}$ & $E_{NOR}$=$E_{gate}$ \\ \hline
        \textbf{OR} & $A_{OR}=1.3A_{gate}$ & $D_{OR}=D_{gate}$ & $E_{OR}=2.3E_{gate}$ \\ \hline
        \textbf{MUX2} & $A_{MUX}=2.2A_{gate}$ & $D_{MUX}=2.2D_{gate}$ & $E_{MUX}=3.0E_{gate}$ \\ \hline
        \textbf{HA} & $A_{HA}=4.3A_{gate}$ & $D_{HA}=2.5D_{gate}$ & $E_{HA}=6.9E_{gate}$ \\ \hline
        \textbf{FA} & $A_{FA}=5.7A_{gate}$ & $D_{FA}=3.3D_{gate}$ & $E_{FA}=8.4E_{gate}$ \\ \hline
        \textbf{DFF} & $A_{DFF}=6.6A_{gate}$ & N/A & $E_{DFF}=9.6E_{gate}$ \\ \hline
        \textbf{SRAM} & $A_{SRAM}=2.2A_{gate}$ & 0 & 0 \\ \hline 
    \end{tabular}

\label{cells cost}
\vspace{-1cm}
}
\end{table}

The number of weights stored in the DCIM array is $W_{store}$. Assuming that the bit width of weight is $B_w$, a total of $(W_{store}*B_w)$ SRAM cells are required to store all the weights. SRAM cells are divided into $N$ columns, and each column contains $(L*H)$ SRAM cells. When weights are stored in a DCIM array, each bit of the weight is mapped to a different column separately. For example, when the weight is 4 bits, 4 bits are mapped to 4 separate columns. Thus, column 1 stores the first bit of the $(L*H)$ weights, column 2 stores the second bit of the $(L*H)$ weights, and so on. To increase the storage density, $L$ weights share 1 Compute Unit. When performing computation, only 1 bit of weight is selected and sent into the Compute Unit. 

Fig.~\ref{Multiply-based} illustrates the detailed Compute Unit design for multiply-based DCIM. For each computation, the selection gate selects a 1-bit weight from the $L$ weights and performs 1-bit × $k$-bit multiplication with the $k$-bit input. Inspired by ~\cite{dcim1}, 1-bit × $k$-bit multiplication is implemented by $k$ 4T NOR gates, where the input of the NOR gate is the $WB$ (inverse of the $W$) and $INB$ (inverse of the $IN$). 

The Adder Tree, consisting of tree-structured adders, is used to sum the outputs of a column of compute cells. Shift Accumulator receives the output from the adder tree. Since the Input Buffer takes $\frac{B_{x}}{k}$ cycles to send the mantissa, the Shift Accumulator needs to collect the partial sums and perform a shift accumulation at each cycle to get the computation results for the full input’s mantissa. Results Fusion Unit is designed on the bottom of the DCIM array to support variable bit-width of weights. Since each column stores only 1 bit of the weight, the results from multiple columns need to be weighted and summed according to the position of the weight bits. 
For example, when the weight is 4 bits, the Results Fusion Unit shifts and sums the results of the 4 columns to obtain the final results $O_{M}$.

The INT-to-FP Converter converts the integer results into FP format. It shifts the long bit-width final result and calculates the exponent and sign bits. Finally, the sign bit, exponent, and mantissa are combined into the FP format. 

\label{section:DCIM performance}

\begin{table*}[tb]
    \centering
    \caption{Symbols and cost model for each component in DCIM.}
    \includegraphics[width=0.92\textwidth]{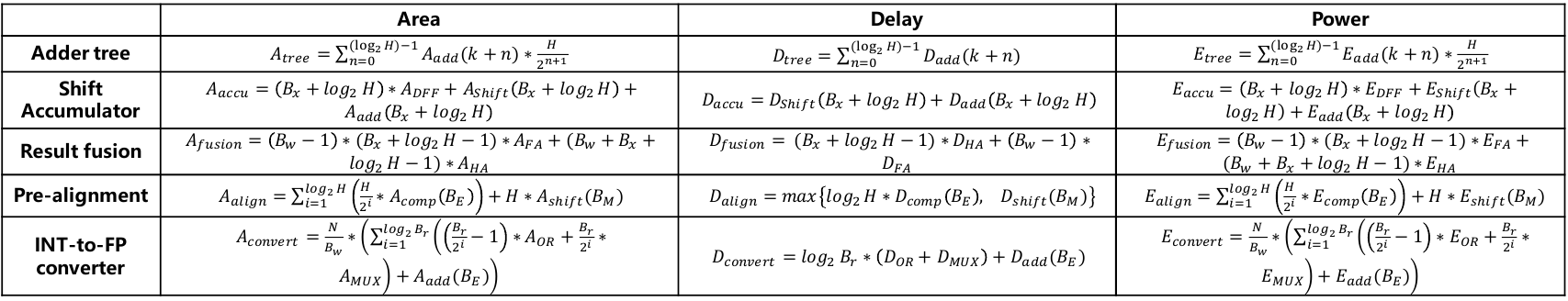}
    \label{component}
    \vspace{-0.3cm}
\end{table*}

\begin{table*}[tb]
    \centering
    \caption{Symbols and cost model for Integer DCIM.}
    \includegraphics[width=0.85\textwidth]{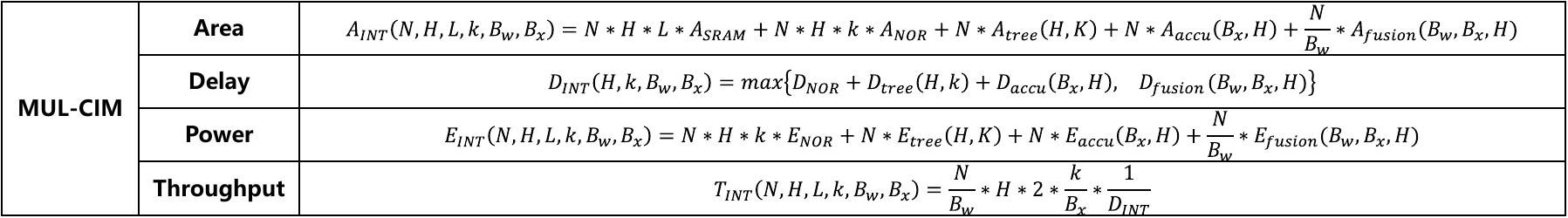}
    \label{INT DCIM}
    \vspace{-0.3cm}
\end{table*}

\begin{table}[tb]
    \centering
    \caption{Symbols and cost model for Float-point DCIM.}
    \includegraphics[width=0.48\textwidth]{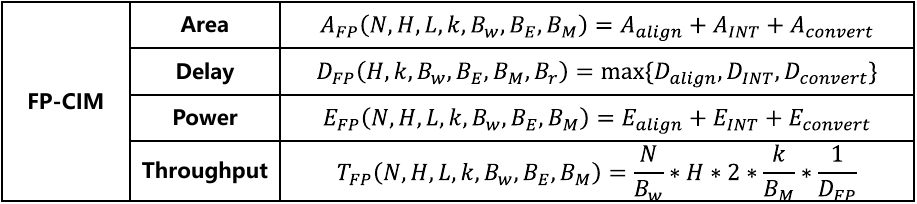}
    \label{FP DCIM}
    \vspace{-0.4cm}
\end{table}

\subsection{MOGA-based Design Space Explorer}
\subsubsection{DCIM Performance Estimation model}

To estimate the area, power consumption, and throughput for SEGA-DCIM, a unified performance estimation model is proposed for DCIM architecture. Since the architecture of DCIM is complex, we start our analysis with digital standard cells. Table \ref{cells cost} shows the area, delay, and power consumption of each standard cell. FA denotes a 1-bit full adder, and HA denotes a 1-bit half adder. For easy of understanding, we normalize all costs to NOR gates based on TSMC28 digital circuits PDK. If the technology process changes, the cost will also be changed. Since the weights are hard-wired from the SRAM cell to the Compute Unit and do not need to be precharged for reading, the latency of the SRAM is 0. Considering that the leakage current of SRAM is very small, the power consumption of the SRAM is approximated to be 0.

Table \ref{module cost} shows the cost of the digital logic modules required in the DCIM. Note that the multiplier in DCIM uses NOR gates for efficient multiplication, as shown in Fig. \ref{Multiply-based}. The $N$-bit adder employs the carry-ripple structure. The $N$-bit shifter utilizes the architecture of a barrel shifter. Thus, an $N$-bit shifter is equal to $N*A_{sel}(N)$. The comparator is only used to select a larger value in the DCIM and is simplified to an $N$-bit adder.

Next, the costs of several DCIM components are analyzed and modeled. The Adder Tree sums the outputs of the Compute Units, typically using a tree-like structure. Assume that there are $H$ inputs to the Adder Tree, and each input is $k$-bits, the area $A_{tree}$, delay $D_{tree}$, and power consumption $E_{tree}$ are shown in Table \ref{component}, respectively.

The Shift Accumulator collects partial sums from the Adder Tree and performs shift accumulation each cycle to obtain the computation results. Assume that the bit-width of input X is $B_{x}$ and the column height is $H$. Since only one bit of weight is stored in each column, as introduced in section III-A, the bit-width of the output of the Shift Accumulator is $(B_{x}+\log _{2} H)$. Thus, it requires $(B_{x}+\log _{2} H)$ registers, one $(B_{x}+\log _{2} H)$-bit shifter, and one $(B_{x}+\log _{2} H)$-bit adder. The area $A_{accu}$, delay $D_{accu}$, and energy $E_{accu}$ are shown in Table \ref{component}, respectively.

The Result Fusion Unit sums the output of the Shift Accumulator based on the position of weight bits to produce the final result for full weight precision. It is necessary to perform a weighted summation of the results from $B_{w}$ columns, and the bit-width of each result is $(B_{x}+\log_{2}H)$-bits. Thus, the area $A_{fusion}$, delay $D_{fusion}$, and power consumption $E_{fusion}$ are shown in Table~\ref{component}, respectively.

After modeling each component of the DCIM, an analysis and modeling of the overall overhead of the DCIM is conducted. For multiply-based integer DCIM, assume the bit-width of the input and weight is $B_{x}$ and $B_{w}$, respectively. The total number of weights stored in the DCIM is $W_{store}=N*H*L/B_{w}$. Thus, the SRAM capacity of the DCIM array is $(N*H*L)$ bits. During computation, the Input Buffer sends $(H*k)$-bits ($1\le k\le B_x$) per cycle to the DCIM array for INT MAC operations. The CIM array also contains $(N*H*k)$ NOR gates, $N$ Adder Trees, $N$ Shift Accumulators, and $\frac{N}{B_{w}}$ Result Fusion Units. Thus, with the aforementioned equation in Table~\ref{component}, the area $A_{INT}$, delay $D_{INT}$, and power consumption $E_{INT}$ are shown in Table \ref{INT DCIM}, respectively.

Since the Shift Accumulator includes registers that implement pipelining, the delay is determined by taking the maximum of two parts. The peak frequency is the reciprocal of the delay. The peak throughput $T_{INT}$ equals the number of operations multiplied by the frequency as shown in Table \ref{INT DCIM}.

For pre-aligned-based FP DCIM, assume that the input's mantissa bit-width is $B_{M}$ and the input's exponent bit-width is $B_{E}$. The FP Pre-alignment module consists of two parts: (1) A set of comparators is used to find the maximum exponent $X_{Emax}$. (2) The subtractor is used to calculate the offset between each exponent and $X_{Emax}$, and the shifter is used to shift the input’s mantissa based on the offset. The area $A_{alig}$, delay $D_{alig}$, and power consumption $E_{alig}$ of the FP Pre-alignment module are shown in Table \ref{component}. The INT-to-FP Converter unit receives the DCIM array final results and converts them to the FP format. The bit-width of the DCIM array output is $B_r=B_w+B_M+{log}_2{H}$. Thus, the area $A_{convert}$, delay $D_{convert}$ and power consumption $E_{convert}$ of the INT-to-FP Converter are shown in Table \ref{component}.

In summary, for the FP DCIM based on pre-aligned, the area $A_{FP}$, delay $D_{FP}$, power consumption $E_{FP}$, and throughput $T_{FP}$ are shown Table \ref{FP DCIM}.

\subsubsection{NSGA-II-based Optimization}

Based on the previous analysis in Section~\ref{section:DCIM performance}, the multi-objective optimization problem of DCIM performance can be formulated as Equation~\ref{finalobjetctint} and Equation~\ref{finalobjetctfloat}, respectively. Equation~\ref{finalobjetctint} demonstrates the situation of multiplier-based DCIMs. The negative sign in front of $f_{-T_{INT}}$ means that it is required to solve for the maximum value. The constraint $k - B_{x} \geq 0$ guarantees that the single-round input bit $k$ can not be larger than the total input. $B_{x}$. $N \cdot H \cdot \frac{L}{B_{w}} = W_{store}$ guarantees the number of storage weights is exactly equal to the user-defined storage weights size. 
\begin{equation}
    \label{finalobjetctint}
    \begin{array}{l@{}l}
        \min_{x} &F(N,H,L,k,B_{w},B_{x})= [A_{INT},D_{INT}, E_{INT}, -T_{INT}] \\
        & \text { s.t. }  k - B_{x} \geq 0 \quad and \quad N \cdot H \cdot \frac{L}{B_{w}} = W_{store}
        % \multicolumn{2}{c}{\text { s.t. }  k - B_{x} \geq 0} \\ 
        % \multicolumn{2}{c}{N \cdot H \cdot \frac{L}{B_{w}} = W_{store}}
    \end{array}
\end{equation}

Equation~\ref{finalobjetctfloat} demonstrates the situation of pre-aligned-based DCIMs with similar constraints as Equation~\ref{finalobjetctint}. After constructing the multi-objective optimization function of multiplier-based and pre-aligned-base DCIMs, a classic NSGA-II algotithm~\cite{pereiraReviewMultiobjectiveOptimization2022} is performed for multiple architectures respectively. Finally, a high-quality Pareto-frontier set containing both integer and floating-point solutions can be obtained with the help of user-defined distillation.

\begin{equation}
    \label{finalobjetctfloat}
    \begin{array}{l@{}l}
         \min_{x} &F(N,H,L,k,B_{w},B_{E},B_{M})= [A_{FP}, D_{FP}, E_{FP}, -T_{FP}] \\
         &\text { s.t. }  k - B_{M} \geq 0 \quad and \quad N \cdot H \cdot \frac{L}{B_{M}} = W_{store}\\
    \end{array}
\end{equation}

\subsection{Template-based DCIM generator}
\begin{sloppypar}
    After obtaining the selected Pareto-frontier DCIMs, the design parameters and architecture are already settled. Based on the template of the specific architecture, the generator will generate the memory array, DCIM compute components, and digital peripherals for the given design parameters, respectively. Finally, all of these components will be merged by a script to generate the final layout.
    
    More specifically, each component generation has two steps, the netlist generation and the layout generation. For the memory array, the netlist and layout are duplicates of fixed computing units. Therefore, the generation follows a fixed rule of duplication and can be easily realized by a script. The users need to give the layout and netlist of basic computing units. For the DCIM compute components and digital peripherals, bit-width is the main change of the DCIM compute components and digital peripherals when switching between different design parameters in a fixed architecture. Therefore the netlist generation process is converted into the Verilog code generation for simplicity. The netlist synthesis process is left to the commercial tools: Innovus. The layout generation of the DCIM compute components and digital peripherals is also conducted by Innovus with predefined constraints. Users can also adjust these constraints directly to meet their customized applications. Finally, the layout can be merged by a script considering the relationship of these three parts. 
\end{sloppypar}

\section{Experimental Results}
\label{sec:Results}

We perform experiments on a Linux server with an Intel Xeon Gold 6248 CPU @ 3.00GHz. SEGA-DCIM is implemented on the TSMC28 PDK with multiple architectures. We conduct experiments on versatile computing precisions, including INT2, INT4, INT8, INT16, FP8, FP16, FP32, and BF16. We also validate SEGA-DCIM with a wide range of $W_{store}$, from 4K to 128K. Among all of these design specifications, the MOGA-based design exploration for a particular array size and computing precision can be finished in 30 minutes. After user-defined distillation, each DCIM design can be generated within one hour thanks to the template-based DCIM generator.  

\begin{figure}[b]
    \vspace{-0.4cm}
    \centering
    \includegraphics[width=0.45\textwidth]{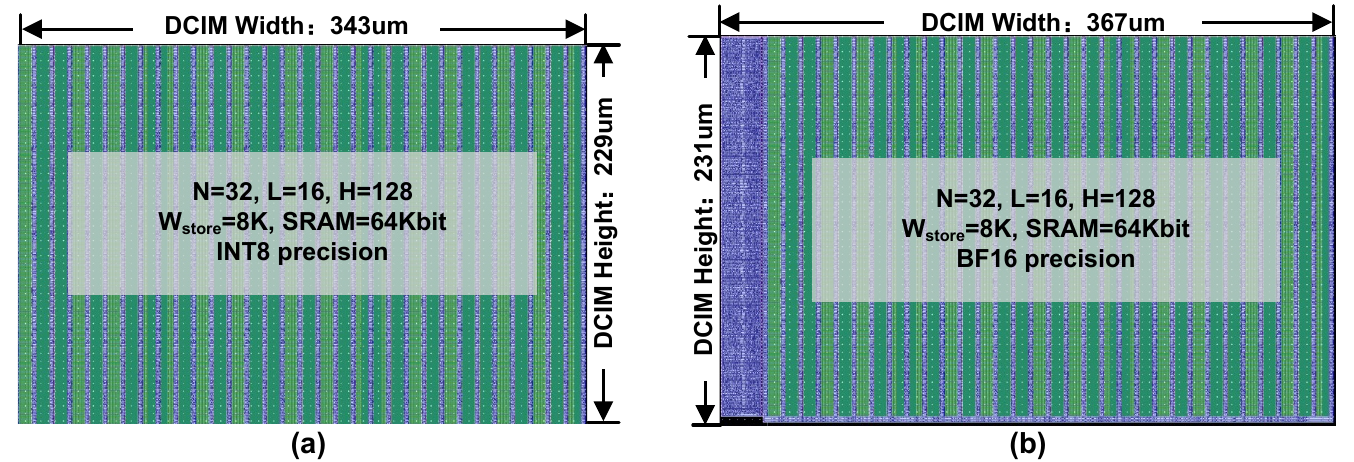}
    \caption{The layouts of DCIM for different data precision.}
    \label{DCIM layout}
    \vspace{-0.3cm}
\end{figure}

Fig.~\ref{DCIM layout} demonstrates the final layout results of DCIM. Both DCIMs contain 8K weights but support different data precision. Fig.~\ref{DCIM layout}(a) shows the multiply-based integer DCIM for INT8 operation. The area of DCIM is 0.079$mm^2$. Fig.~\ref{DCIM layout}(b) shows the pre-aligned-based floating-point DCIM for BF16 operations. The area of DCIM is 0.085$mm^2$ while the area of pre-aligned-based circuits is only 0.006$mm^2$, demonstrating the high area efficiency for floating-point operation. 

\begin{figure}[tb]
    \centering
    \includegraphics[width=0.45\textwidth]{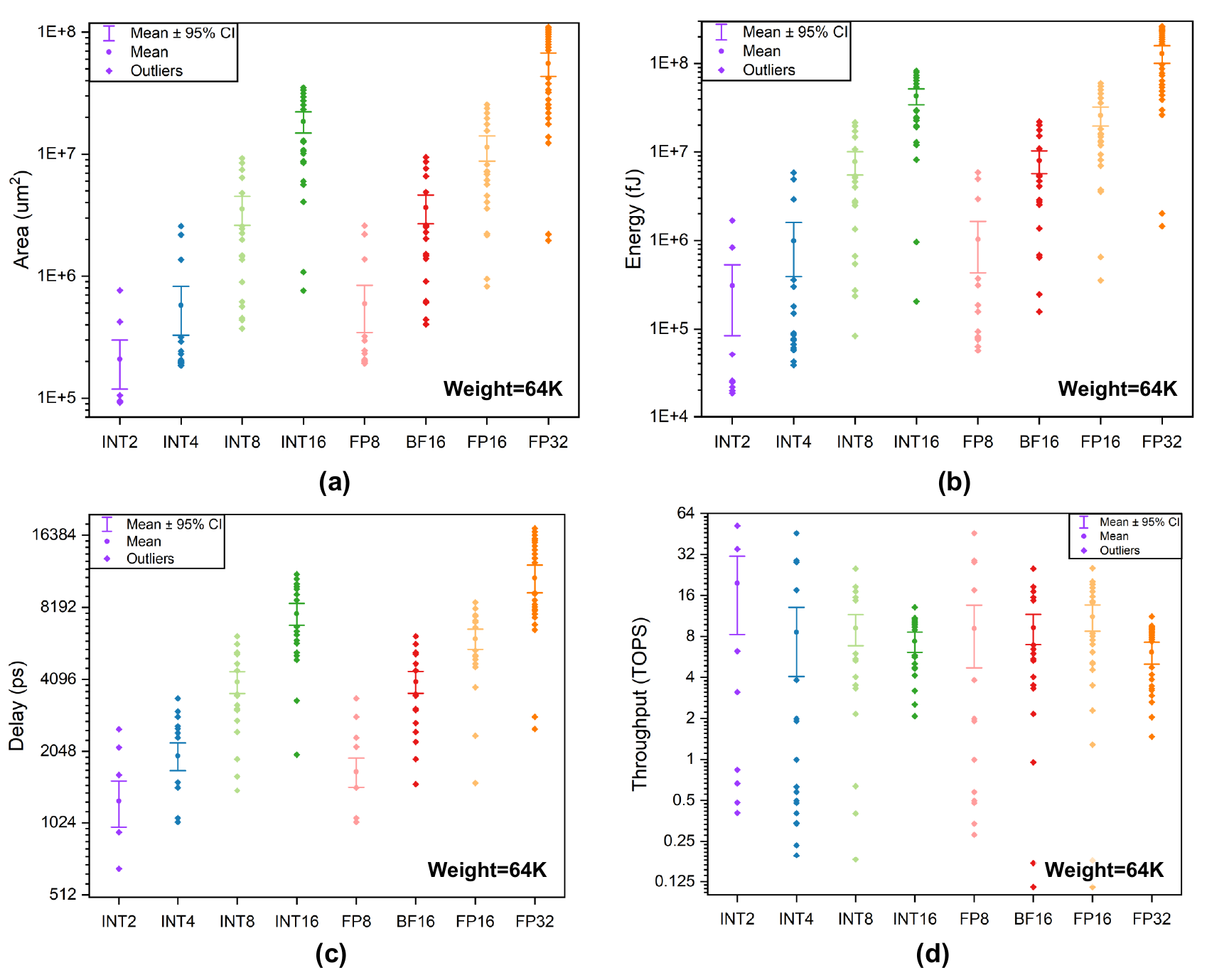}
    \caption{ Design Space of SEGA-DCIM with different precision.}
    \label{ Design Space of SEGA-DCIM}
    \vspace{-0.5cm}
\end{figure}

A comprehensive analysis of SEGA-DCIM design space is performed by a MOGA-based design space explorer. To avoid generating extreme results, $N$ is set to be greater than $4*B_w$, $L$ is set to be no greater than 64, and $H$ is set to be no greater than 2048 during the design space exploration. Fig.~\ref{ Design Space of SEGA-DCIM} shows the design space of SEGA-DCIM. To make it easier to understand, we only show the results with $W_{store}=64K$ and analyze the effect of different data precision on the results. Fig.~\ref{ Design Space of SEGA-DCIM}(a)(b)(c)(d) presents the trend of area, energy, delay, and throughput under different data precision. As the data precision grows from INT2 to FP32, the average area grows from 0.2$mm^2$ to 60$mm^2$, the average energy grows from 0.3$nJ$ to 103$nJ$, and the average delay grows from 1.2$ns$ to 10.9$ns$. Note that the overhead for floating-point precision does not increase much compared to integer precision. For example, the overhead of BF16 is almost the same compared to INT8, demonstrating the efficiency of the pre-aligned architecture for performing floating-point operations. 

% Figure~\ref{ Design Space of SEGA-DCIM}(e)(f) presents the trend of the energy efficiency and area efficiency under different data precision. Energy efficiency is measured by $TOPS/W$ (Tera Operations per Second per Watt) at 0.9V supply voltage and 10\% sparsity. This metric presents the trade-off between throughput and energy. When the data precision is increased from INT2 to FP32, the energy will be higher for the same throughput, leading to an energy efficiency drop from 303 $TOPS/W$ to 1.2 $TOPS/W$. Area efficiency is measured by $TOPS/mm^2$ (Tera Operations per Second per Square Millimetre) at 0.9V supply voltage. This metric presents the trade-off between throughput and area. When data precision is increased from INT2 to FP32, more digital logic circuits are required for the same throughput, leading to an area efficiency drop from 106 $TOPS/mm^2$ to 0.02 $TOPS/mm^2$.

\begin{figure}[tb]
    \centering
    \includegraphics[width=0.45\textwidth]{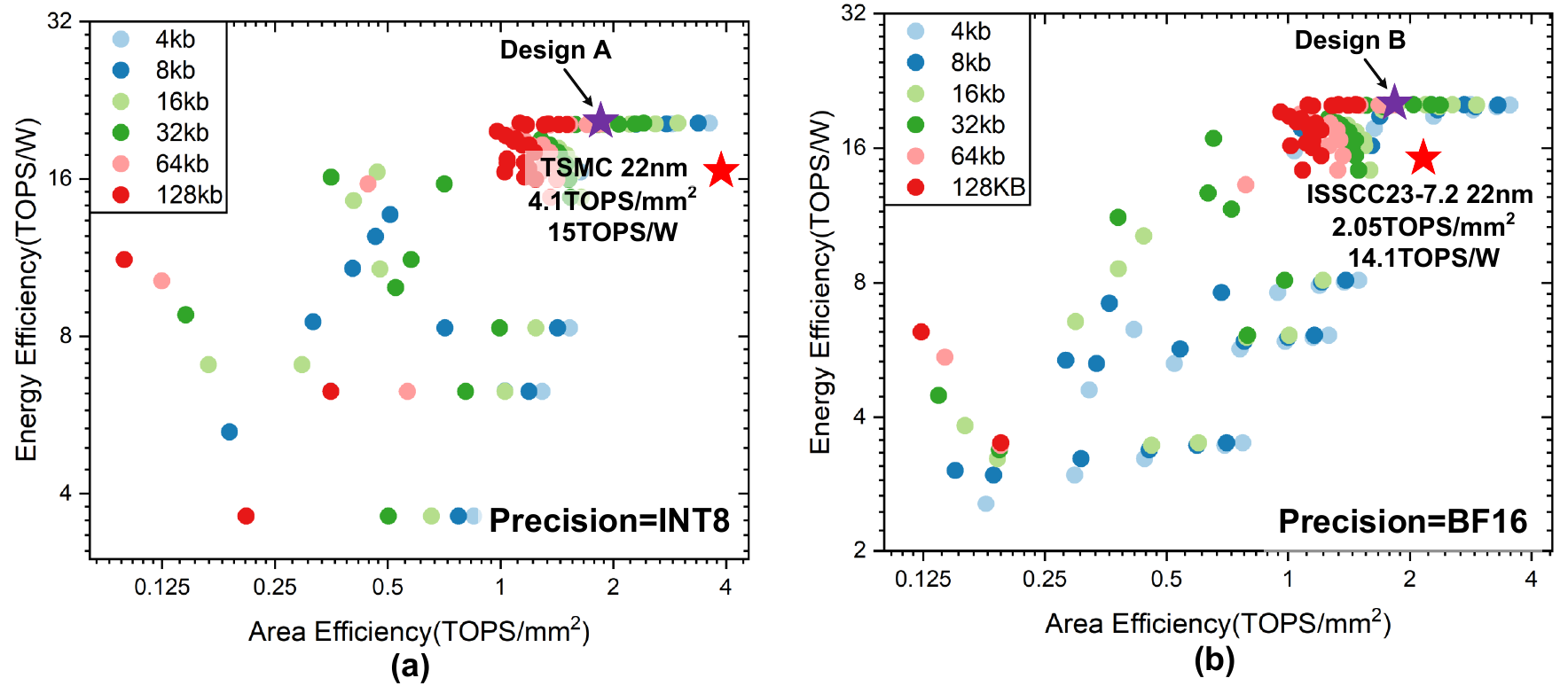}
    \caption{Comparison between SEGA-DCIM and SOTA works.}
    \label{result Comparasion}
    \vspace{-0.5cm}
\end{figure}

Fig.~\ref{result Comparasion} shows the energy efficiency and area efficiency comparison between SEGA-DCIM and SOTA DCIM works. Energy efficiency is measured by $TOPS/W$ (Tera Operations per Second per Watt) at 0.9V supply voltage and 10\% sparsity. Area efficiency is measured by $TOPS/mm^2$ (Tera Operations per Second per Square Millimetre) at 0.9V supply voltage. The data precision is fixed and the $W_{store}$ size is changed. Fig.~\ref{result Comparasion}(a) shows the results for INT8 data precision. We add the TSMC's recent DCIM work~\cite{dcim1}, which contains 64K weights and fabricated in a 22nm technology process, into figure~\ref{result Comparasion}(a). For a fair comparison, we chose design A with 64K weights and achieved 22$TOPS/W$ energy efficiency and 1.9$TOPS/mm^2$ area efficiency. Our design achieves a higher energy efficiency but with a lower area efficiency than TSMC's work. This is because TSMC uses foundry SRAM arrays to reduce the area of SRAM. Fig.~\ref{result Comparasion}(b) shows the results for BF16 data precision. We add the recent ISSCC-23's work~\cite{dcim_float2}. It contains 64K weights and is fabricated in a 22nm technology process. We chose design B with 64K weights and achieved 20.2$TOPS/W$ energy efficiency and 1.8$TOPS/mm^2$ area efficiency. Same as in Fig.~\ref{result Comparasion}(a), our design achieves a higher energy efficiency, but with a lower area efficiency than ISSCC's work.

\section{Conclusion}
\label{sec:Conclusion}

\begin{sloppypar}
In this paper, we propose SEGA-DCIM, a design space exploration-guided automatic digital CIM compiler. With the MOGA-based design space explorer and estimation model, SEGA-DCIM can automatically generate high-quality DCIM designs for versatile application scenarios. SEGA-DCIM can obtain the Pareto-frontier of the DCIM designs with a user-defined number of weights and computing precision. The DCIM solutions have a wide design space with computing precision ranging from INT2 to FP32. Validated in TSMC28, SEGA-DCIM can generate DCIM solutions with competitive performance compared with SOTA DCIMs designed by hand, and the experimental results demonstrate the robustness and benefits of SEGA-DCIM.

\end{sloppypar}

\bibliographystyle{IEEEtran}
\bibliography{./ref/ICCAD24}

\begin{thebibliography}{10}
\providecommand{\url}[1]{#1}
\csname url@samestyle\endcsname
\providecommand{\newblock}{\relax}
\providecommand{\bibinfo}[2]{#2}
\providecommand{\BIBentrySTDinterwordspacing}{\spaceskip=0pt\relax}
\providecommand{\BIBentryALTinterwordstretchfactor}{4}
\providecommand{\BIBentryALTinterwordspacing}{\spaceskip=\fontdimen2\font plus
\BIBentryALTinterwordstretchfactor\fontdimen3\font minus \fontdimen4\font\relax}
\providecommand{\BIBforeignlanguage}[2]{{%
\expandafter\ifx\csname l@#1\endcsname\relax
\typeout{** WARNING: IEEEtran.bst: No hyphenation pattern has been}%
\typeout{** loaded for the language `#1'. Using the pattern for}%
\typeout{** the default language instead.}%
\else
\language=\csname l@#1\endcsname
\fi
#2}}
\providecommand{\BIBdecl}{\relax}
\BIBdecl

\bibitem{Jia-JSSC}
H.~Jia, M.~Ozatay, Y.~Tang, H.~Valavi, R.~Pathak, J.~Lee, and N.~Verma, ``Scalable and programmable neural network inference accelerator based on in-memory computing,'' \emph{IEEE Journal of Solid-State Circuits}, vol.~57, no.~1, pp. 198--211, 2022.

\bibitem{Song-eDRAMCIM}
J.~Song, X.~Tang, H.~Luo, H.~Zhang, X.~Qiao, Z.~Sun, X.~Yang, Z.~Wu, Y.~Wang, R.~Wang, and R.~Huang, ``A 4-bit calibration-free computing-in-memory macro with 3t1c current-programed dynamic-cascode multi-level-cell edram,'' \emph{IEEE Journal of Solid-State Circuits}, vol.~59, no.~3, pp. 842--854, 2024.

\bibitem{sunAnalogDigitalInmemory}
J.~Sun, P.~Houshmand, and M.~Verhelst, ``Analog or {{Digital In-memory Computing}}? {{Benchmarking}} through {{Quantitative Modeling}},'' \emph{ICCAD 2023}.

\bibitem{Wang-PositCIM}
Y.~Wang, X.~Yang, Y.~Qin, Z.~Zhao, R.~Guo, Z.~Yue, H.~Han, S.~Wei, Y.~Hu, and S.~Yin, ``34.1 a 28nm 83.23tflops/w posit-based compute-in-memory macro for high-accuracy ai applications,'' in \emph{2024 IEEE International Solid-State Circuits Conference (ISSCC)}, vol.~67, 2024, pp. 566--568.

\bibitem{dcim1}
Y.-D. Chih, P.-H. Lee, H.~Fujiwara, Y.-C. Shih, C.-F. Lee, R.~Naous, Y.-L. Chen, C.-P. Lo, C.-H. Lu, H.~Mori, W.-C. Zhao, D.~Sun, M.~E. Sinangil, Y.-H. Chen, T.-L. Chou, K.~Akarvardar, H.-J. Liao, Y.~Wang, M.-F. Chang, and T.-Y.~J. Chang, ``16.4 an 89tops/w and 16.3tops/mm2 all-digital sram-based full-precision compute-in memory macro in 22nm for machine-learning edge applications,'' in \emph{2021 IEEE International Solid- State Circuits Conference (ISSCC)}, vol.~64, 2021, pp. 252--254.

\bibitem{jssc-hkdiao}
H.~Diao, Y.~He, X.~Li, C.~Tang, W.~Jia, J.~Yue, H.~Luo, J.~Song, X.~Li, H.~Yang, H.~Jia, Y.~Liu, Y.~Wang, and X.~Tang, ``A multiply-less approximate sram compute-in-memory macro for neural-network inference,'' \emph{IEEE Journal of Solid-State Circuits}, pp. 1--12, 2024.

\bibitem{dcim_float2}
A.~Guo, X.~Si, X.~Chen, F.~Dong, X.~Pu, D.~Li, Y.~Zhou, L.~Ren, Y.~Xue, X.~Dong, H.~Gao, Y.~Zhang, J.~Zhang, Y.~Kong, T.~Xiong, B.~Wang, H.~Cai, W.~Shan, and J.~Yang, ``A 28nm 64-kb 31.6-tflops/w digital-domain floating-point-computing-unit and double-bit 6t-sram computing-in-memory macro for floating-point cnns,'' in \emph{2023 IEEE International Solid-State Circuits Conference (ISSCC)}, 2023, pp. 128--130.

\bibitem{Yue_FPCIM}
J.~Yue, C.~He, Z.~Wang, Z.~Cong, Y.~He, M.~Zhou, W.~Sun, X.~Li, C.~Dou, F.~Zhang, H.~Yang, Y.~Liu, and M.~Liu, ``A 28nm 16.9-300tops/w computing-in-memory processor supporting floating-point nn inference/training with intensive-cim sparse-digital architecture,'' in \emph{2023 IEEE International Solid-State Circuits Conference (ISSCC)}, 2023, pp. 1--3.

\bibitem{dcim_float3}
H.~Diao, H.~Luo, J.~Song, B.~Xu, R.~Wang, Y.~Wang, and X.~Tang, ``A 28nm 128tflops/w computing-in-memory engine supporting one-shot floating-point nn inference and on-device fine-tuning for edge ai,'' in \emph{2024 IEEE Custom Integrated Circuits Conference (CICC)}, 2024, pp. 1--2.

\bibitem{chenNeuroSimCircuitLevelMacro2018}
P.-Y. Chen, X.~Peng, and S.~Yu, ``{{NeuroSim}}: {{A Circuit-Level Macro Model}} for {{Benchmarking Neuro-Inspired Architectures}} in {{Online Learning}},'' \emph{IEEE Transactions on Computer-Aided Design of Integrated Circuits and Systems}, vol.~37, no.~12, pp. 3067--3080.

\bibitem{neurosim}
X.~Peng, S.~Huang, Y.~Luo, X.~Sun, and S.~Yu, ``Dnn+neurosim: An end-to-end benchmarking framework for compute-in-memory accelerators with versatile device technologies,'' in \emph{2019 IEEE International Electron Devices Meeting (IEDM)}, 2019, pp. 32.5.1--32.5.4.

\bibitem{gonugond2020}
S.~K. Gonugondla, C.~Sakr, H.~Dbouk, and N.~R. Shanbhag, ``Fundamental {{Limits}} on {{Energy-Delay-Accuracy}} of {{In-Memory Architectures}} in {{Inference Applications}},'' \emph{IEEE Transactions on Computer-Aided Design of Integrated Circuits and Systems}, vol.~41, no.~10, pp. 3188--3201.

\bibitem{kamineniMemGenOpenSourceFramework2021}
S.~Kamineni, S.~Gupta, and B.~H. Calhoun, ``{{MemGen}}: {{An Open-Source Framework}} for {{Autonomous Generation}} of {{Memory Macros}},'' in \emph{2021 {{IEEE Custom Integrated Circuits Conference}} ({{CICC}})}.\hskip 1em plus 0.5em minus 0.4em\relax {IEEE}, pp. 1--2.

\bibitem{liuOpenSAROpenSource2021}
M.~Liu, X.~Tang, K.~Zhu, H.~Chen, N.~Sun, and D.~Z. Pan, ``{{OpenSAR}}: {{An Open Source Automated End-to-end SAR ADC Compiler}},'' in \emph{2021 {{IEEE}}/{{ACM International Conference On Computer Aided Design}} ({{ICCAD}})}.\hskip 1em plus 0.5em minus 0.4em\relax {IEEE}, pp. 1--9.

\bibitem{zhangEasyACIM}
H.~Zhang, J.~Song, X.~Gao, X.~Tang, Y.~Lin, R.~Wang, and R.~Huang, ``Easyacim: An end-to-end automated analog cim with synthesizable architecture and agile design space exploration,'' in \emph{Proceedings of the 61st ACM/IEEE Design Automation Conference}, ser. DAC '24, New York, NY, USA, 2024.

\bibitem{chenAutoDCIMAutomatedDigital}
J.~Chen, F.~Tu, K.~Shao, F.~Tian, X.~Huo, C.-Y. Tsui, and K.-T. Cheng, ``Autodcim: An automated digital cim compiler,'' in \emph{2023 60th ACM/IEEE Design Automation Conference (DAC)}, 2023, pp. 1--6.

\bibitem{pereiraReviewMultiobjectiveOptimization2022}
J.~L.~J. Pereira, G.~A. Oliver, M.~B. Francisco, S.~S. Cunha, and G.~F. Gomes, ``A {{Review}} of {{Multi-objective Optimization}}: {{Methods}} and {{Algorithms}} in {{Mechanical Engineering Problems}},'' \emph{Archives of Computational Methods in Engineering}, vol.~29, no.~4, pp. 2285--2308.

\bibitem{sainiMultiobjectiveOptimizationTechniques2021}
N.~Saini and S.~Saha, ``Multi-objective optimization techniques: A survey of the state-of-the-art and applications: {{Multi-objective}} optimization techniques,'' \emph{The European Physical Journal Special Topics}, vol. 230, no.~10, pp. 2319--2335.

\bibitem{ngatchouParetoMultiObjective2005}
P.~Ngatchou, A.~Zarei, and A.~El-Sharkawi, ``Pareto {{Multi Objective Optimization}},'' in \emph{Proceedings of the 13th {{International Conference}} on, {{Intelligent Systems Application}} to {{Power Systems}}}.\hskip 1em plus 0.5em minus 0.4em\relax {IEEE}, pp. 84--91.

\end{thebibliography}
\vfill
\end{document}